\documentclass[a4paper,twocolumn,11pt,accepted=2025-10-19]{quantumarticle}
\pdfoutput=1
\usepackage[utf8]{inputenc}
\usepackage[T1]{fontenc}
\usepackage{graphicx}
\usepackage{dcolumn}
\usepackage{bm}
\usepackage{braket}
\usepackage{multirow}
\usepackage{hhline}
\usepackage{mathtools}
\usepackage{csquotes}
\usepackage{bookmark}
\usepackage{hyperref}
\begin{document}
\title{LUCI in the surface code with dropouts}

\newcommand{\goog}{\affiliation{Google Quantum AI, Santa Barbara, CA 93117, USA}}
\newcommand{\googla}{\affiliation{Google Quantum AI, Venice, CA 90291, USA}}
\newcommand{\googjp}{\affiliation{Google Quantum AI, Tokyo, Japan}}

\author{Dripto M. Debroy}
\email{dripto@google.com}
\googla

\author{Matt McEwen}
\goog

\author{Craig Gidney}
\goog

\author{Noah Shutty}
\googla

\author{Adam Zalcman}
\googla
\googjp

\begin{abstract}
Recently, usage of detecting regions facilitated the discovery of new circuits for fault-tolerantly implementing the surface code. 
Building on these ideas, we present LUCI, a framework for constructing fault-tolerant circuits flexible enough to construct aperiodic and anisotropic circuits, making it a clear step towards quantum error correction beyond static codes. 
We show that LUCI can be used to adapt surface code circuits to lattices with imperfect qubit and coupler yield, a key challenge for fault-tolerant quantum computers using solid-state architectures.
These circuits preserve spacelike distance for isolated broken couplers or isolated broken measure qubits in exchange for halving timelike distance, substantially reducing the penalty for dropout compared to the state of the art and creating opportunities in device architecture design.
For qubit and coupler dropout rates of $1\%$ and a patch diameter of 15, LUCI achieves an average spacelike distance of 13.1, compared to 9.1 for the best method in the literature. 
For a SI1000(0.001) circuit noise model, this translates to a $36\times$ improvement in median logical error rate per round, a factor which increases with device performance.
At these dropout and error rates, LUCI requires roughly $25\%$ fewer physical qubits to reach algorithmically relevant one-in-a-trillion logical codeblock error rates.

\end{abstract}

\maketitle

\section{Introduction}\label{sec:intro}
In order to reach the error rates necessary for large-scale quantum algorithms, we will require the use of quantum error correction (QEC). By most estimates, such a computer would require thousands of logical qubits, each composed of hundreds to thousands of physical qubits~\cite{burg2021quantum, lee2021even, kim2022fault, ShokrianZini2023quantumsimulationof, rubin2023fault}. In solid-state architectures, fabrication errors can lead to a number of failure modes that result in broken qubits, as well as broken couplers in architectures that use them, like Google's Sycamore and USTC's Zuchongzhi~\cite{arute2019quantum, wu2021strong, zhu2022quantum}. 

In addition, transient errors can significantly degrade qubit or coupler performance for long time periods, such as drifting Two-Level Systems (TLSs)~\cite{klimov2018fluctuations}.
\begin{figure}[t!]
    \centering    \includegraphics[width=.99\linewidth]{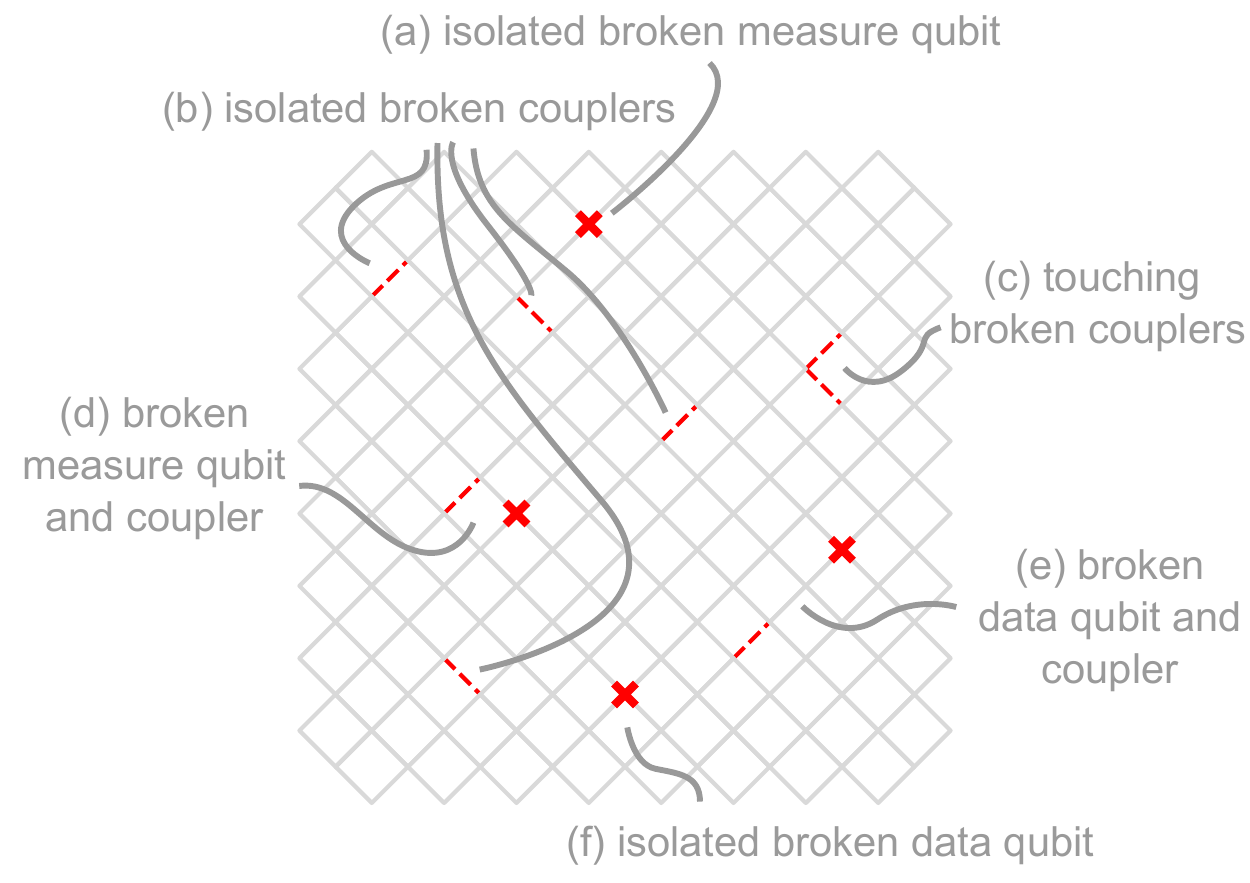}
    \caption{
    \textbf{An example dropout grid. }
    A set of dropouts on a distance-9 surface code chip with 169 qubits and 360 couplers. Qubits are nodes and couplers are edges, with broken components shown in red. There are a number of important configurations: (a) isolated broken measurement qubit, (b) isolated broken couplers, (c) two broken couplers on the same qubit, (d) broken measure qubit next to broken coupler, (e) broken data qubit near broken coupler such that the affected data qubits are across a measure qubit, and (f) broken data qubit. (c) and (d) are special cases in LUCI, while (e) is a case that is handled differently between the two methods we use as comparisons.}
    \label{fig:example grid}
\end{figure}
We refer to qubits or couplers that are either permanently or temporarily unusable in the context of a QEC protocol as \emph{dropouts}. 
\begin{figure*}[t!]
    \centering
    \includegraphics[width=0.97\linewidth]{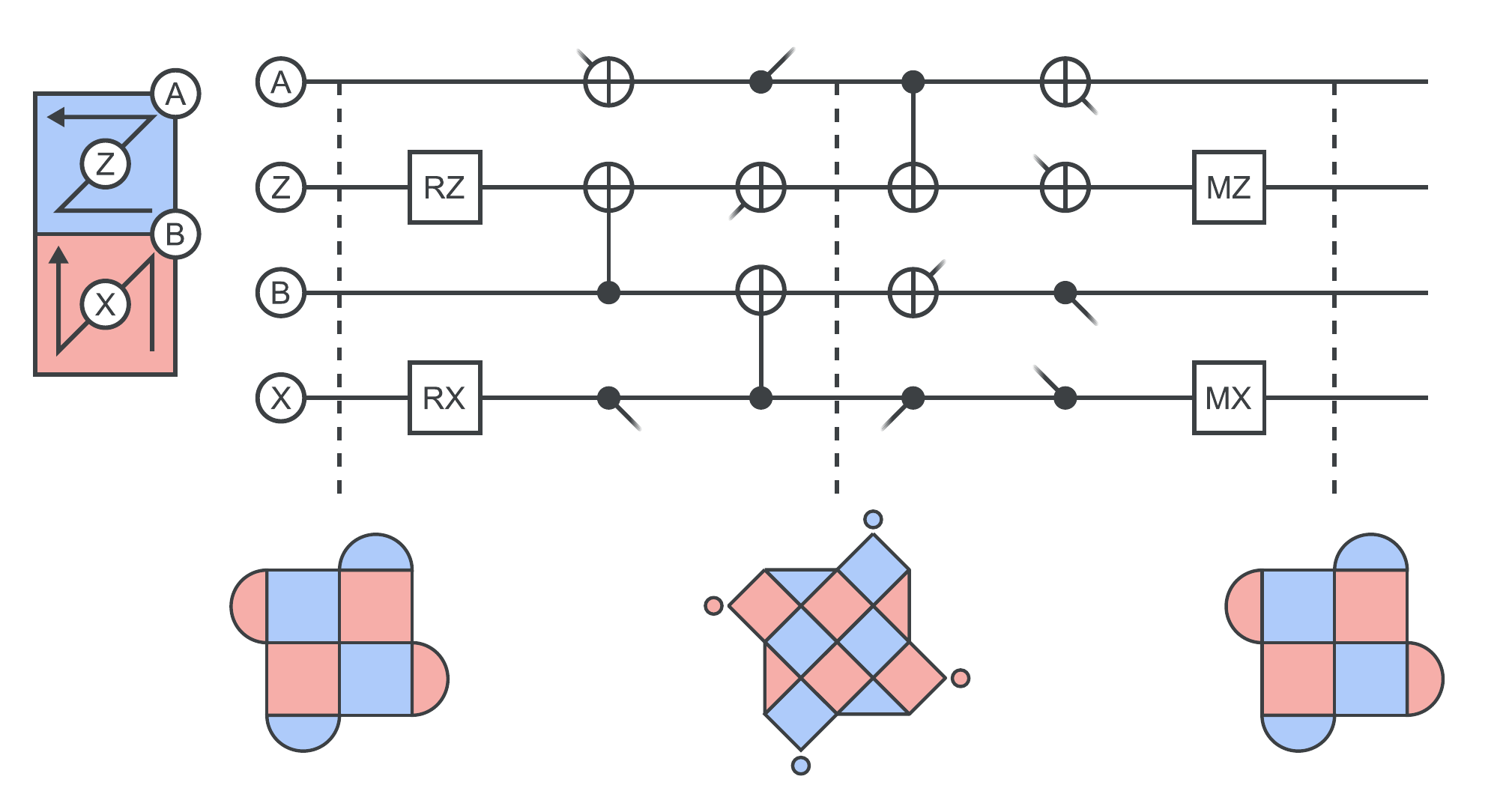}
    \caption{
    \textbf{Detection region slices in the Surface Code. }
    Adapted from \protect\cite{mcewen2023relaxing}. (top) Zoomed in view of the circuit used to measure a standard surface code, with angled CNOT gates used to indicate gates which connect to qubits not labeled. (bottom) Timelike slices of the detecting regions of the surface code circuit at the indicated point. Crucially, the mid-cycle state of surface code circuit is an unrotated surface code. In this figure and throughout this manuscript, blue(red) will be used to indicate Z(X)-type Pauli operators unless specified otherwise. For visual clarity, weight-1 operators are represented as circles slightly offset from the relevant qubit.}
    \label{fig:sc detslice}
\end{figure*}
In this manuscript we will focus on a simplified failure model, where qubits and couplers are missing from the grid with independent probabilities $p_q$ and $p_c$, respectively. In Fig.~\ref{fig:example grid} we show an example dropout grid for a chip designed to fit a distance-9 surface code, where around 2\% of the qubits and couplers are missing. 

There are a number of methods known for adapting quantum error correction protocols to grids with qubit and coupler dropout~\cite{auger2017fault, strikis2023quantum, siegel2023adaptive, lin2024codesign, wei2024low, GransSamuelsson2024improvedpairwise}. In most cases, isolated broken data qubits reduce distance by one in both directions, isolated broken measure qubits reduce distance by two. Isolated broken couplers are fixed by removing the data qubit which interacts via that coupler. More substantial differences between the methods arise when dealing with multiple broken components in a small area, with the method described in \cite{auger2017fault}, and later rediscovered by \cite{wei2024low}, better preserving the functional parts of the chip.

In this manuscript we will discuss a new method for building quantum error correction circuits for grids with dropouts, based on some of the ideas first introduced in \cite{mcewen2023relaxing}. These circuits are constructed from a small set of different \emph{rounds}, each starting and ending in a modified mid-cycle state of the surface code. 
We introduce a visual language for indicating which mid-cycle stabilizers are measured in that round, and via which operations. Individual mid-cycle stabilizer measurements are described by \emph{shapes}, including L-, U-, C-, and I-shapes, amongst others, so these diagrams are referred to as LUCI diagrams. 

\begin{figure*}[ht!]
    \centering
    \includegraphics[width=0.97\linewidth]{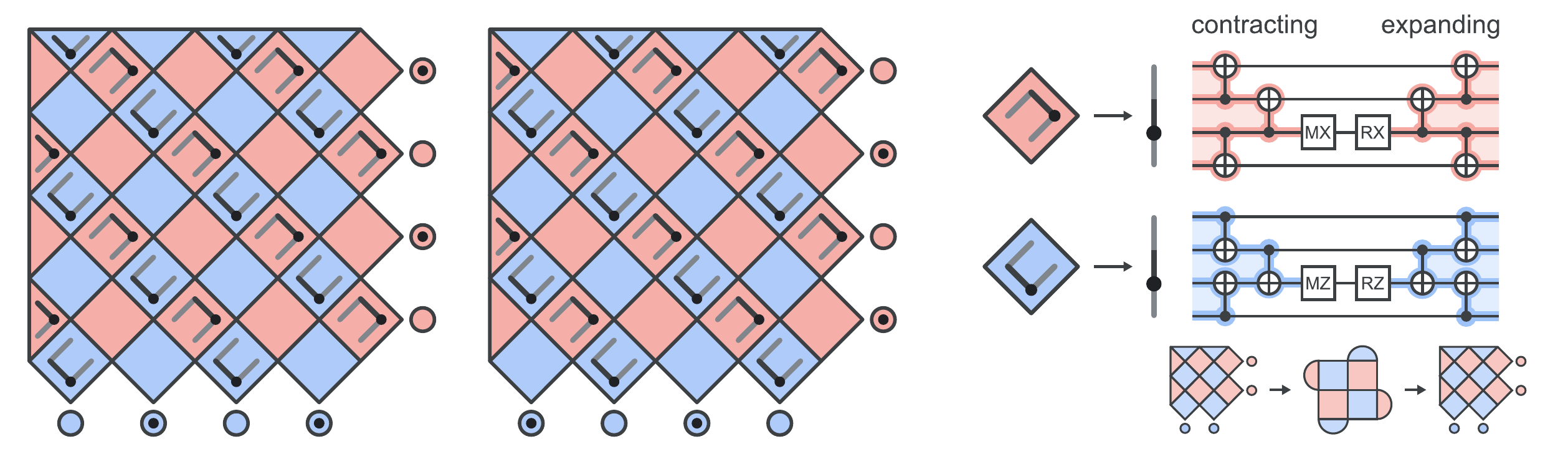}
    \caption{
    \textbf{LUCI Diagrams and their circuit interpretation. }
    (left) LUCI diagram showing the two rounds of a distance-5 three-coupler surface code circuit~\protect\cite{mcewen2023relaxing}.
    Colors indicate the mid-cycle stabilizers of the surface code being measured, with X stabilizers in red and Z stabilizers in blue. The gray shapes inside the squares indicate how that stabilizer will be measured. (right) Circuit compilations for U-shapes on X-type (red) and Z-type (blue) squares. The colored regions depict the detecting region \emph{contracting} and being measured out, and then a new detecting region \emph{expanding} from a reset. Shades of gray in the shape denote which circuit layer the operation occurs in. Other shapes, like the L's on the top and left boundaries, are formed by removing the appropriate CNOT gates from these U-shape compilations.}
    \label{fig:LUCI diagrams}
\end{figure*}
We provide an algorithm which constructs a valid LUCI diagram given a qubit grid with dropouts.
Circuits built using this technique far outperform the best known techniques in the literature in spacelike distance, at the expense of halving the timelike distance relative to other dropout methods. Here, we use spacelike distance to refer to the minimum length of a logical operator crossing from one spatial boundary to the other in a memory experiment, and timelike distance to refer to the length of a logical operator connecting two timelike boundaries in a stability experiment or a lattice surgery operation. The circuits produced achieve this spacelike improvement by using available components more efficiently to work around damaged areas, and consequently far outperform previous methods in terms of logical error rate. 

\section{LUCI}\label{sec:LUCI}
This section is organised as follows: First, we discuss the underpinnings of the LUCI framework, approaching circuit construction from the mid-cycle state of the surface code. 
Next, we define LUCI diagrams and their interpretation as circuits, as well as explaining the constraints necessary to make valid diagrams.
Finally, we work through an algorithm for constructing valid LUCI diagrams given a specific set of dropouts, eventually handling the large example introduced in Fig.~\ref{fig:example grid}.

\subsection{Detecting regions and the mid-cycle state}\label{sec:mid-cycle}
In Ref.~\cite{mcewen2023relaxing}, the authors introduce an approach to error correction circuits where the standard stabilizers of the quantum error correcting code are propagated through the circuit to form so-called ``detecting regions". 
Detecting regions capture the extent of the stabilizer in spacetime, where Pauli errors are detectable by looking at a specific set of measurement outcomes. 
By approaching the problem of QEC circuit construction as the task of covering the circuit with appropriate detecting regions (as opposed to constructing the circuit directly from a code), we reveal additional freedom to manipulate the circuit and improve performance.

Cross-sections of the detecting regions at key points during a standard surface code circuit are shown in Fig.~\ref{fig:sc detslice}. In the bulk rounds of the circuit detecting regions survive for two rounds, starting at measure qubit initialization, expanding into the typical code stabilizer in the first round of entangling gates, and then contracting back to be terminated in the second round. Detectors can be formed by combining all the measurements that a given detecting region terminates on, which for the standard circuit is simply a comparison of subsequent measurements on the relevant measure qubit. In LUCI circuits, we produce more complex structures of detecting regions that span additional rounds and terminate on spatially separate measurements. As such, the detecting region picture is particularly important in understanding and building the detectors for our circuits, which is described in Sec.~\ref{sec:diagram to circuit}.

One insight of Ref.~\cite{mcewen2023relaxing} is that the mid-cycle state of the standard surface code state is an unrotated surface code state on both measure and data qubits. This allowed the construction of novel surface code circuits by measuring the mid-cycle stabilizers in different ways before returning to the same state. In this view, the circuit is constructed from mid-cycle state to mid-cycle state, with half-rounds at the beginning and end of the circuit to get to the usual initial and final states of a surface code circuit. 
LUCI approaches the task of constructing circuits around dropouts in a similar way, using the mid-cycle surface code state as a ``home base" to return to. 

The LUCI construction in particular has similarities to the three-coupler surface code circuit from Ref.~\cite{mcewen2023relaxing}. Using a pair of CNOT gates, a weight-4 bulk mid-cycle stabilizer is ``folded'' onto the pair of qubits along one edge of the square. This weight-2 operator is then folded again using a single CNOT gate, after which it is measured and reset, before being unfolded back into the original weight-4 footprint. Adjacent mid-cycle stabilizers can be measured simultaneously using a ``snake'' configuration, where the gates are positioned such that the CNOT gates for the initial circuit layer fold both stabilizers simultaneously. This snake configuration will be seen again in Sec.~\ref{sec:LUCI diagrams}, as it will be the rule we use to fit LUCI shapes together correctly.

\subsection{LUCI diagrams}\label{sec:LUCI diagrams}
LUCI diagrams provide a visual language for describing circuit constructions starting at the mid-cycle state. 
In order to construct a LUCI circuit, we start by finding a set of mid-cycle stabilizer generators which are compatible with the dropout grid in question. Some of these stabilizers may be composed of multiple mid-cycle gauge operators multiplied together, as in other subsystem code dropout constructions. 
We then specify a set of shapes that describe how to measure these mid-cycle gauge operators and stabilizers over multiple rounds.
Each of the rounds from mid-cycle to mid-cycle is described by a board in the LUCI diagram, an example of which is shown in Fig.~\ref{fig:LUCI diagrams}. 
A LUCI circuit iterates through these boards sequentially, returning to the mid-cycle state as a reliable interface between rounds. 
Because the mid-cycle state is stable, the process of finding logical operators simply requires finding an Pauli string in the mid-cycle state from one corner to another, and then propagating it through the circuit.
The detector cross-section at measurement varies for each round, as illustrated in Appendix~\ref{app:detslices}, indicating that LUCI circuits implement a dynamic code. 

On the left side of Fig.~\ref{fig:LUCI diagrams}, we present a LUCI diagram describing the two rounds of the distance-5 three-coupler surface code circuit. Each round measures some subset of the mid-cycle stabilizers of the code. On the right side, we show how these shapes can be interpreted as quantum circuits, with diagrams of the mid-cycle state at key points. In the contracting stage of the round, CNOT gates propagate the information to a single qubit and measure it. Then, in the expanding stage of the round, the same qubit is reset, and then the CNOT gates are repeated in reverse order to spread the information back to the original footprint. As long as each mid-cycle stabilizer is measured in one of the rounds, the full distance of the code will be achieved. 

\begin{figure}[t!]
    \centering
    \includegraphics[width=0.9\linewidth]{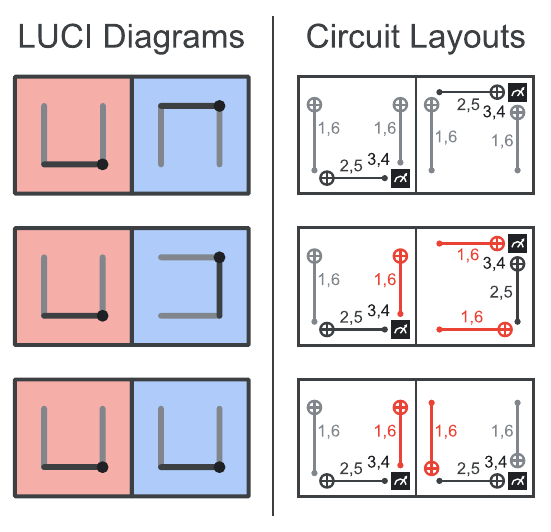}
    \caption{
    \textbf{Compatibility of neighboring LUCI shapes. }
    Examples of adjacent squares which are compatible (top row) and incompatible (bottom two rows). Like for the LUCI diagrams, qubits in the circuit layouts are at the vertices of the square grid, with gate layers indicated by the numbers to match with the compilation in Fig.~\protect\ref{fig:LUCI diagrams}. The dial icon indicates a MRX or MRZ gate for red or blue squares, respectively. For the two incompatible examples, the gates highlighted in red show the collision which makes the diagram invalid.}
    \label{fig:shape compatibility}
\end{figure}
\begin{figure}[b!]
    \centering
    \includegraphics[width=0.95\linewidth]{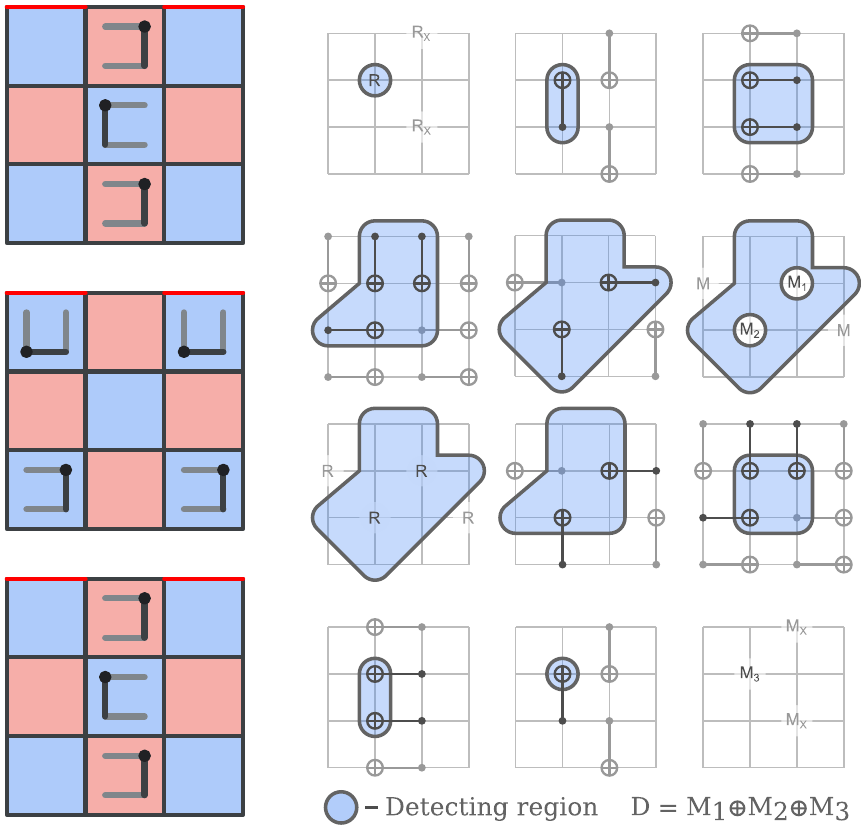}
    \caption{
    \textbf{Construction of a detecting region from shapes. }
    Timelike slices of a single Z-type detecting region, starting from the reset layer of the first round (top row), before being pulled in multiple direction and passing through two measures and resets during the second round (middle two rows), and then finally being measured out completely in the first half of the third round (bottom row). Gates are dark gray if they non-trivially modify the detecting region, which for a CNOT means the incoming Z-type Pauli operator must have support on the target of the CNOT gate. Measurements are in the Z basis unless indicated otherwise.}
    \label{fig:detecting region}
\end{figure}
The dark gray shapes in the diagrams fully describe how the mid-cycle stabilizer will be measured, with the lines indicating which entangling operations are used and the dot indicating which qubit is measured, as shown on the right side of Fig.~\ref{fig:LUCI diagrams}. As a result, we must use compatibility rules for adjacent shapes to ensure that the resulting circuits never have a gate collision, where two distinct gates use a shared qubit in the same round. This means that the layer-1 CNOT gate between the shared qubits must be identical, and the layer-2 CNOT gates on the shared qubits must not overlap. These constraints lead to the ``snake'' pattern shown in the LUCI diagram on the left of the figure, where a full diagonal of mid-cycle stabilizers is measured simultaneously using shared layer-1 CNOTs and alternating layer-2 CNOTs. A few examples are presented in Fig.~\ref{fig:shape compatibility} to explain this further. The only cases where these shapes do not give enough information is for L-shapes with the measure qubit in the middle, where the order of CNOTs is ambiguous, and I-shapes, where the layer the CNOT is applied is ambiguous. In both cases, one can infer what is happening by seeing how the adjacent squares are measured and using the compatibility rules.

An interesting note is that in a LUCI diagram, reversing the orientation of the second layer of CNOTs allows you to switch the qubits that are measured without impacting compatibility. As a result, it is straightforward to build LUCI circuits which switch data qubit and measure qubit roles in successive rounds, which can be impactful for leakage errors~\cite{mcewen2023relaxing, camps2024leakagemobilitysuperconductingqubits}. One could do this within a usual four-round cycle, or double the cycle to eight round and switch qubit roles for the second half of the set of eight. This flexibility is one of the primary advantages of LUCI, making it easy to modify a circuit to measure particularly leaky qubits more often than others without much difficulty, or whatever else suits the experimental conditions.

\subsection{Generating Circuits from LUCI Diagrams}\label{sec:diagram to circuit}
One method to guarantee all squares are measured when modifying diagrams is to use a four-coloring of the underlying square lattice of the mid-cycle state, shown in the seventh step of Fig.~\ref{fig:LUCI example2}. Since the squares of a given color do not share any qubits, they can all be filled independently without any compatibility concerns, in this way LUCI allows us to measure shapes which would otherwise be incompatible by interleaving in time. By requiring that each square is filled in its highlighted round, we guarantee a fully measured mid-cycle state. Additional squares can also be measured in each round to minimize logical error rate, and in Sec.~\ref{sec:LUCI example} we build around a base diagram, to both maximize the number of measurements and make the circuits more structured, reducing calibration overhead. This method guarantees a valid LUCI diagram for arbitrary configurations of dropouts.

Once a valid diagram has been generated, it can be used to create circuits.
To initialize, we use just the ``expanding'' half, as labeled in Fig.~\ref{fig:LUCI diagrams}, of the first round, with the reset layer filled in to reset all other qubits in the logical initialization basis.
The subsequent two CNOT layers get us into the mid-cycle state.
We then cycle between the different rounds, appending one less than the desired number of total rounds, to account for the half rounds on each end of the circuit.
To measure, we apply only the ``contracting'' half of the final round, with the CNOTs moving us back to the end-cycle, and apply a measurement in the desired logical measurement basis to all the qubits which would normally be left unmeasured in the round.
This gives the operations for a full memory experiment.

\begin{figure}[t!]
    \centering
    \includegraphics[width=0.9\linewidth]{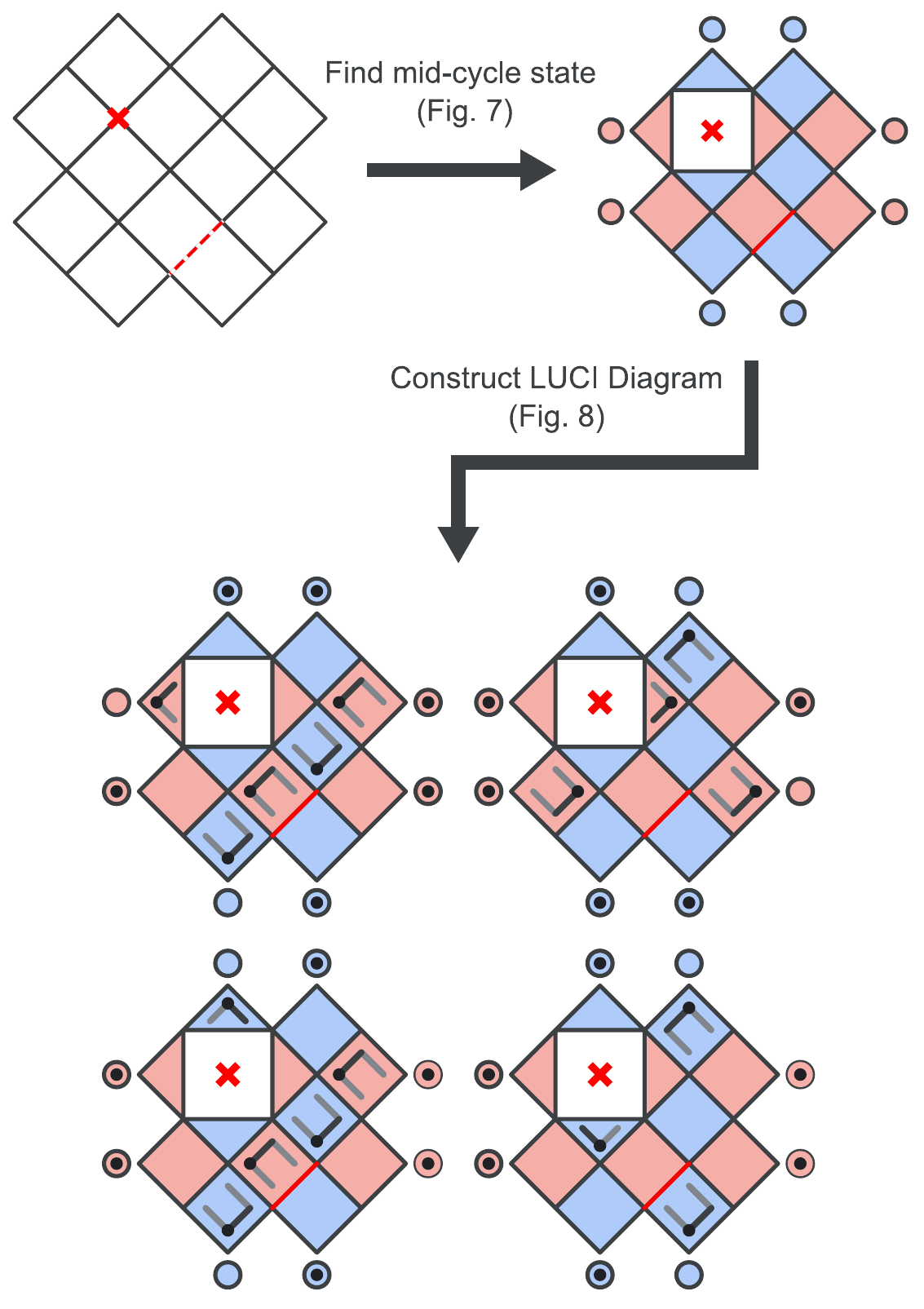}
    \caption{\textbf{Overview of LUCI Algorithm. }The first step, described in detail in Fig.~\protect\ref{fig:LUCI example1} and the accompanying caption, finds the appropriate mid-cycle state to build around for the dropout grid. In this case the broken coupler does not require modification, while the broken qubit is removed from the support of the nearby operators. The second step involves actually constructing a valid LUCI diagram for the mid-cycle state of interest, and is described in Fig.~\protect\ref{fig:LUCI example2} and the accompanying caption.}
    \label{fig:alexis}
\end{figure}
We show the structure for a representative detecting region in Fig.~\ref{fig:detecting region}, For LUCI, we have to be more careful than in the standard surface code case. Unlike in the usual circuit, boards avoiding dropouts may pull detecting regions outwards before returning them into their usual position, leading to detectors that consist of a number of measurements on different qubits, as can be seen in other middle-out constructions~\cite{mcewen2023relaxing, shaw2024loweringconnectivityrequirementsbivariate}. A detecting region starts at a reset, after which it takes two CNOT layers to expand into a mid-cycle stabilizer. The next round generally does not measure this same mid-cycle stabilizer, and in this case the detecting region may be manipulated by the measurement of nearby mid-cycle stabilizers. The region is returned to its original position, possibly by including measurements and resets on the way. It then is folded and measured fully, completing the region. The corresponding detector combines all the measurements which touch the region. 

\subsection{Building an example LUCI circuit}\label{sec:LUCI example}
\begin{figure*}
    \centering
    \includegraphics[width=0.95\linewidth]{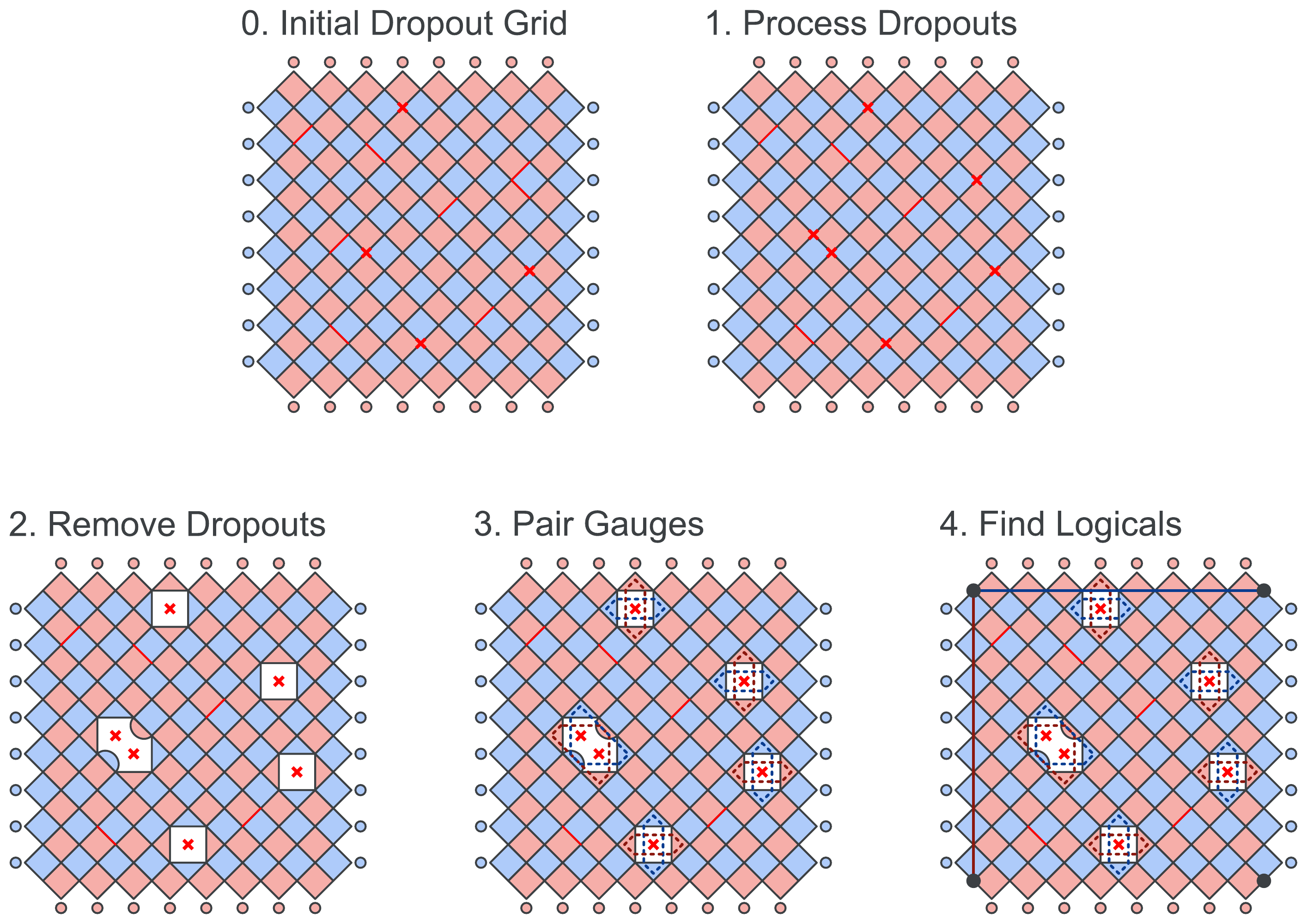}
    \caption[]{
    \textbf{Finding the mid-cycle state.}\hspace{\textwidth}\textbf{1. }Starting from the original dropout grid in Fig~\protect\ref{fig:example grid}, convert each broken qubit into four broken couplers connected to that qubit, then remove any qubit which has two perpendicular broken couplers attached to it. LUCI handles broken couplers by passing information around the broken coupler, which is impossible if an additional qubit or coupler along that path is broken. As a result, qubits isolated by multiple broken components in a single mid-cycle stabilizer are removed. A qubit is removed on the left due to a broken coupler being too close to a broken qubit, and one is removed from the top right for having two incident broken couplers.\hspace{\textwidth}\textbf{2. }Remove all broken qubits from mid-cycle stabilizers they touch. This will lead to some non-commuting Pauli operators, which we will refer to as \textit{mid-cycle gauge operators}. Like in a subsystem code, the gauge operators commute with all stabilizers, but may anti-commute with other gauge operators. All remaining broken couplers do not prevent us from measuring the usual stabilizers, so they are not treated differently at this stage.\hspace{\textwidth}\textbf{3. }Form super-stabilizers by grouping mid-cycle gauge operators into products which commute with all other gauge operators. Extending stabilizers into detecting regions keeps commutation relations, so doing this grouping at the mid-cycle gives the same results as grouping at the end-cycle. This grouping problem admits multiple solutions, so we choose the groupings that minimize super-stabilizer size. In the figure, grouped mid-cycle gauge operators are indicated by the red and blue dotted regions. As explained in \protect\cite{auger2017fault}, this process can be thought of as combining stabilizer generators of the same type that touch a given broken qubit, forming new stabilizers which are not supported on the qubit in question. Broken components that are clustered or near the boundaries may lead to gauge operators which cannot be paired, and must be removed or handled differently. This is discussed further in App.~\protect\ref{app:boundary dropouts}.\hspace{\textwidth}\textbf{4. }Finding mid-cycle logical operators. We start by identifying the corners of the code, qubits which are in a single stabilizer of each type. In the figure these qubits are indicated by dark gray circles. Bare logical X(Z) operators, shown in red(blue), can be found by connecting two corners on opposite X(Z) boundaries of the mid-cycle state with operators that commute with every mid-cycle stabilizer and mid-cycle gauge operator. In the detecting region picture, these logical operators are sheets in the 2+1-dimensional view of the circuit, so identifying the logical operator at the mid-cycle fully defines the operator at all other time slices. Over the course of the circuit the logical operator will move due to CNOT gates, but return to the same support for each mid-cycle state.
    }
    \label{fig:LUCI example1}
\end{figure*}
\begin{figure*}
    \centering
    \includegraphics[width=0.90\linewidth]{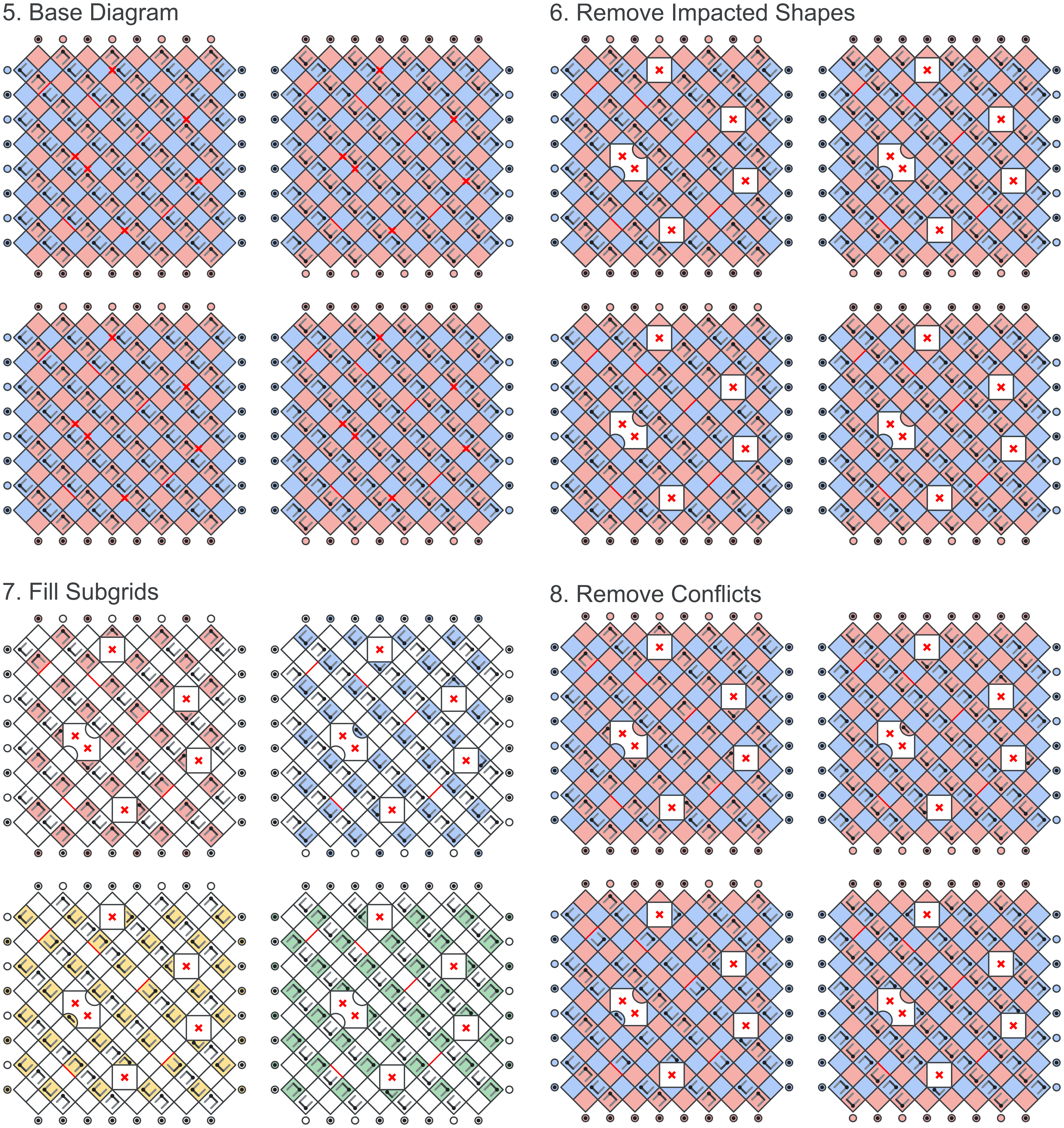}
    \caption[]{
    \textbf{Constructing a LUCI Diagram.}\hspace{\textwidth}\textbf{5. }Start with a base LUCI diagram for a 4-coupler surface code. This circuit is nearly identical to the usual surface code, except that the CNOT order for the stabilizer measurements is reversed every other round and the boundaries are filled in with measure qubits, as opposed to alternating like the usual circuit.\hspace{\textwidth}\textbf{6. }Swap in the updated mid-cycle state found in step 2, removing any shapes which touch broken components or otherwise conflict with the new mid-cycle state. Some shapes will already be compatible with the missing couplers, like the bottom left broken coupler in round 4 in the figure.\hspace{\textwidth}\textbf{7. }Apply a four-coloring to the mid-cycle state with each square colored in only one round and no overlapping qubits between squares of the same color. Insert shapes in the gaps created in step 6 for rounds where the empty square is colored. the color constraint guarantees every square is measured over the four rounds, while the spacing of the colored squares guarantees that we never have conflicts between squares of the same color. Note that we select a four-coloring where the stabilizers and gauge operators highlighted in the first two rounds are Z-type and the second two are X-type, allowing us to combine gauge operators into super-stabilizers before their eigenvalues are scrambled by the anti-commuting gauge operators of the other basis.\hspace{\textwidth}\textbf{8. }The previous step may have introduced incompatibilities. Resolve conflicts by removing shapes not on their colored squares. At this stage additional post-processing and optimization can also be done.
    }
    \label{fig:LUCI example2}
\end{figure*}
In Fig.~\ref{fig:alexis} we provide a simplified example of the process of generating a LUCI diagram for a small dropout grid. Once the diagram is created, we can compile it into a circuit as described in the previous sections. In Figs.~\ref{fig:LUCI example1} and \ref{fig:LUCI example2}, we go through the steps of building a LUCI diagram for the larger dropout grid in Fig.~\ref{fig:example grid}. Steps 1-4 involve finding appropriate mid-cycle stabilizers for the dropout grid, followed by Steps 5-8 where a valid LUCI diagram is constructed for the grid in question. In App.~\ref{app:boundary dropouts} we discuss the case where there are dropouts on the boundary, and other cases where super-stabilizers must be merged. There are other niche cases which can be dealt with more carefully, like parallel broken qubits on the same mid-cycle stabilizer. The goal of this section is to understand the general idea of how to build LUCI circuits, and we leave such special cases to the reader.

The described method is simply one way to build a LUCI diagram for a dropout grid. There are a number of optimizations that can be made on top of this method to target different error models and hardware constraints. As an example, more measurements can be inserted by flipping the orientations of entire rows to alleviate incompatibility issues. However, shapes with are oriented opposite to each other lead to situations in which detecting regions of one type are stretched beyond the usual size, while the other basis is unstretched. This produces uneven detection event fractions, causing logical error rate biases and worsened performance when decoding. Additional optimizations include removing boundary qubits which are only ever used by single-qubit mid-cycle stabilizers and as edge qubits in weight-4 stabilizers, as the weight-4 stabilizer could be converted into an L-shape without damaging the code. This would reduce the footprint of the code without a penalty on performance, depending on the error model. In this paper we will keep these extra qubits to avoid additional complexity.
\begin{figure}[t!]
    \centering
    \includegraphics[width=0.6\linewidth]{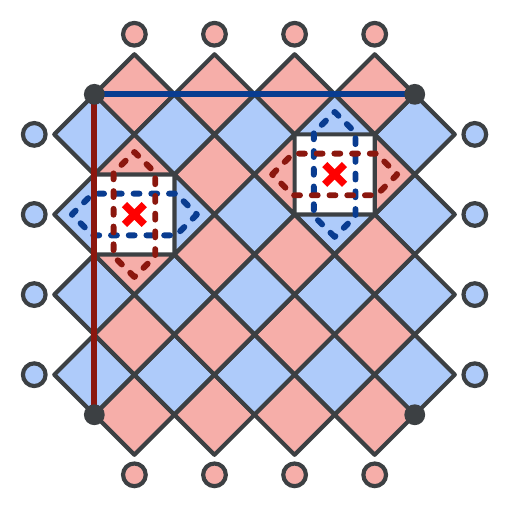}
    \caption{\textbf{Explaining distance loss from qubit dropouts. } The mid-cycle state for a grid with two qubit dropouts. The right side dropout creates Z(X)-type stabilizers which are extended vertically(horizontally). This means that the X(Z)-type logical operator, shown in red(blue), can use the extended stabilizer as a shortcut to reduce the logical distance by one. In contrast, the left side dropout has stabilizers extended parallel to the logical operator of their own type, so they do not reduce distance.}
    \label{fig:dq dropout}
\end{figure}

For intuition as to why this method outperforms the currently known state of the art, we point to two key features. Firstly, the holes that are cut around broken components are much smaller than in previous methods since they exist in the mid-cycle state. This means that we do not have as many issues with nearby holes merging together, and suffer less of a performance penalty from individual dropouts. The super-stabilizers that are formed also are oriented such that they do not have additional extent along logical operators for measure qubits. In Fig.~\ref{fig:dq dropout}, we show the orientations of super-stabilizers in the mid-cycle state for a grid with two missing qubits. It can be seen that one qubit has super-stabilizers formed such that the X(Z)-type super-stabilizer is elongated in the direction of the Z(X)-type logical operator, reducing distance in both cases, while the other qubit has the super-stabilizers elongated perpendicular to the relevant logical operator. While we explain this behavior using the language of data and measure qubits for simplicity, the cause of this behavior is geometric, and independent on whether one compiles a circuit that would measure the missing qubit or not.

In addition, LUCI circuits can be generated by randomly sampling arbitrary applicable shapes in the subgrids of step 7. The subgrid framework guarantees that the detecting regions only span 4 rounds, so this method could be used to create aperiodic and anisotropic error correcting circuits which still maintain spacelike distance and perform reasonably close to the usual surface code. Such circuits may not be relevant for most platforms, but are interesting in terms of vastly expanding the space of circuits usable for error correction, and could be useful for random compiling~\cite{beale2023randomizedcompilingfaulttolerantquantum, jain2024improved}.

\section{Results}\label{sec:Results}
\begin{figure}[b!]
    \centering
    \includegraphics[width=0.97\linewidth]{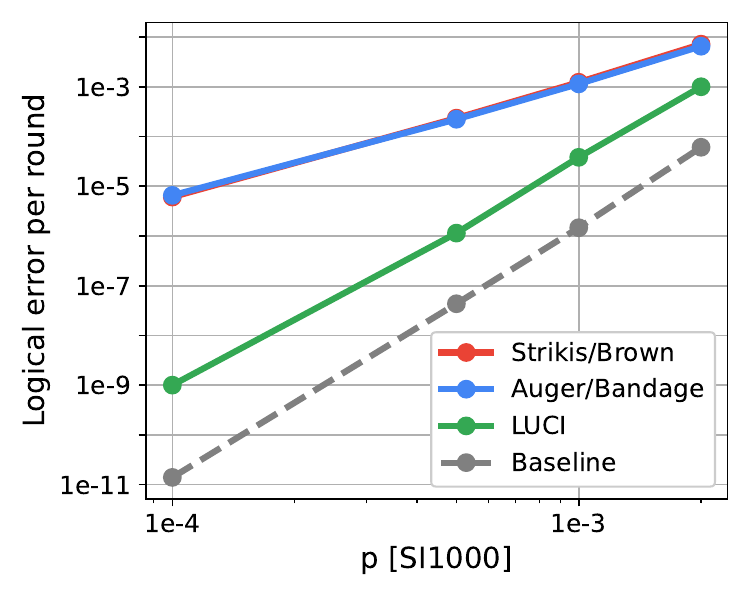}
    \caption{\textbf{Logical error rate versus physical error rate for example grid. }A plot showing logical error rate per round versus SI1000 error rate, for the Strikis/Brown method (red, under blue curve), the Auger/Bandage method (blue), and LUCI (green), for the $\ell=9$ example dropout grid shown in Fig.~\protect\ref{fig:example grid}. A dropout-free distance-9 surface code (dotted gray) is shown as a reference.}
    \label{fig:example grid simulation}
\end{figure}
\begin{figure*}[ht!]
    \centering
    \includegraphics[height=50ex]{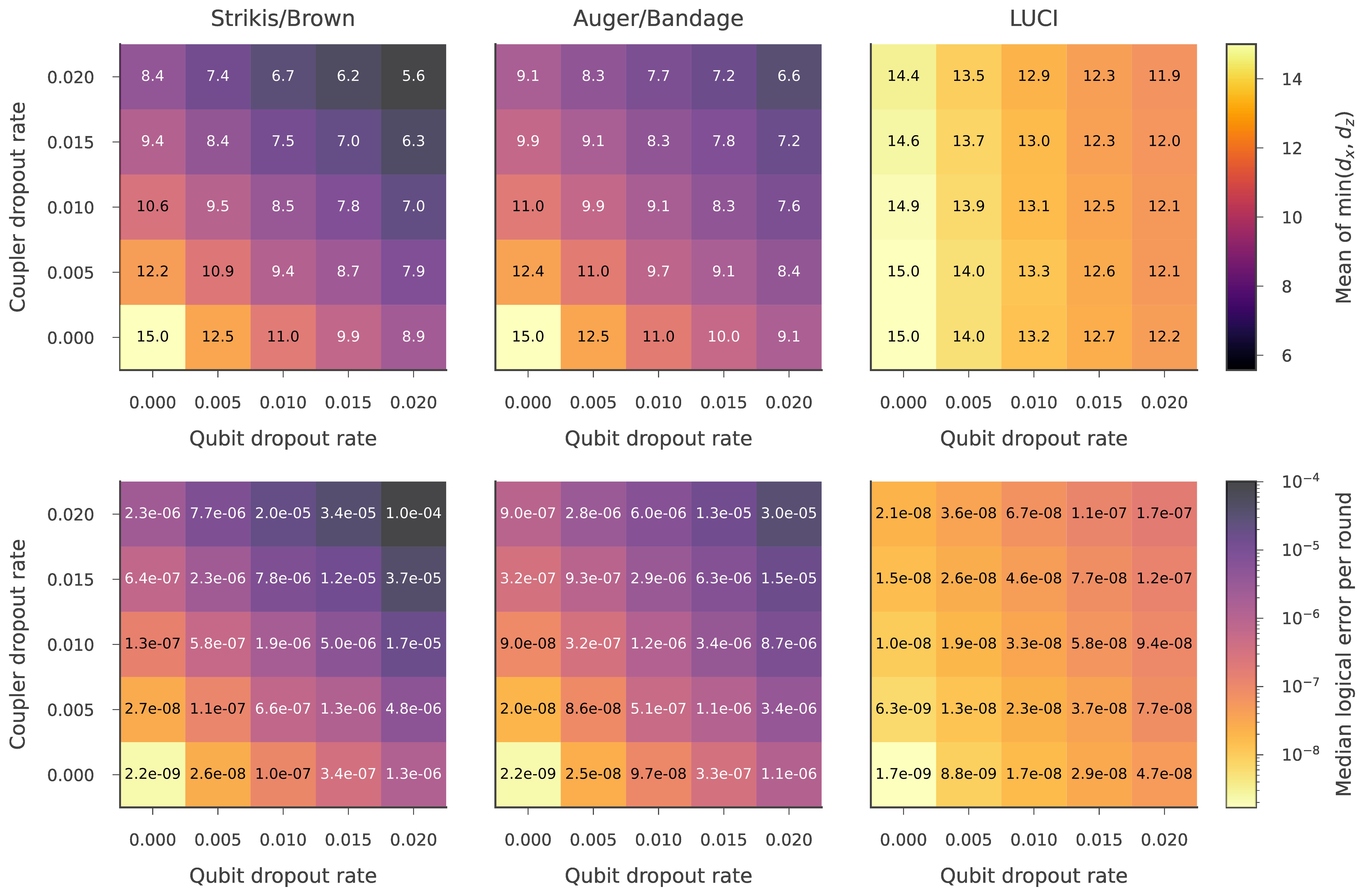}\includegraphics[height=50ex]{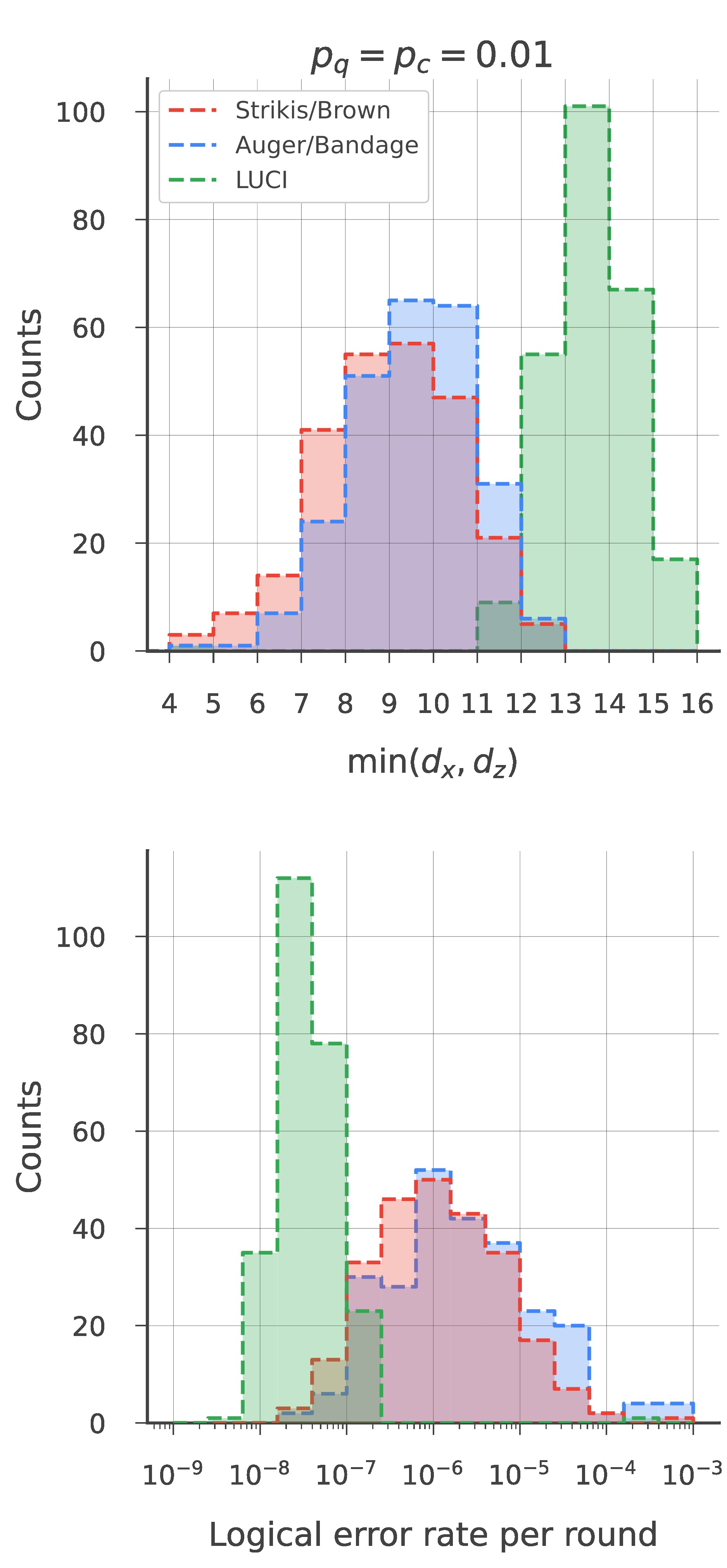}
    \caption{\textbf{Comparison of methods at scale. }Heatmaps for all three methods showing achieved distance averaged over both bases (top), and logical error rate per round (bottom) averaged over 250 randomly sampled $\ell = 15$ rotated surface code grids with the given dropout parameters. LUCI performs the best in both categories, and shows a clear asymmetry between broken couplers and qubits due to LUCI not losing spacelike distance for isolated broken couplers. The other two methods, which remove the attached data qubit for a broken coupler and consequently lose distance, do not see as strong of an asymmetry. On the right side we present histograms for the ensemble of $p_q = p_c = 0.01$ grids, showing far tighter distributions for LUCI relative to the other methods. All simulations are for SI1000(0.001) noise and the same grid instances were used for each method.}
    \label{fig:heatmaps}
\end{figure*}
In this section we will compare LUCI to the two other methods of handling dropout mentioned in the introduction. We will refer to the method from \cite{auger2017fault} and \cite{wei2024low} as the Auger/Bandage method, and the method from \cite{strikis2023quantum, siegel2023adaptive, lin2024codesign} as the Strikis/Brown method. As shown in \cite{wei2024low}, the Strikis/Brown method removes additional qubits and is strictly worse than the Auger/Bandage method, but is helpful because the single-type boundaries make certain proofs more straightforward. We keep it as a comparison point since it is well known, and as a result serves as a point of reference. We use \texttt{stim} for all simulations, with a correlated minimum-weight perfect matching decoder used for decoding and SI1000 noise~\cite{gidney2021fault}, which sets error rates for different operations based on a single parameter $p$, described in Appendix~\ref{app:si1000}. We simulate X and Z logical memory experiments for $\ell \times \ell \times 4\ell$ code blocks, where $\ell$ is the patch diameter of the grid, and convert the resulting average logical error to a per-round logical error rate. For LUCI circuits we define a round as a single round of the LUCI diagram, meaning that we will use each round $\ell$ times over the course of the $4\ell$ rounds. As these different methods have different timelike distance, this comparison is slightly complicated, as depending on the timelike distance of your circuit you may need more rounds. LUCI circuits only promise to measure all stabilizers over four rounds, while the other two methods measure all stabilizers in two. We note, however, that the LUCI circuits still measure the majority of stabilizers at the same rate as standard methods, and only at low error rates will the sparse low-weight logical operators dominate performance. In practice, one should run stability experiments to chose the number of rounds in a logical idle block~\cite{Gidney2022stability}. In the results presented in Figs.~\ref{fig:example grid simulation} and \ref{fig:heatmaps}, we report per-round logical error rates, and a conservative reader may wish to multiply the logical error rate per round numbers by a factor of two to be consistent with the timelike distance of the other methods. The circuits used are provided on Zenodo\cite{zenodo}.

For the Auger/Bandage method and LUCI, we use custom written circuit generation tools, while for the Strikis/Brown method we use the circuit generation software helpfully provided in \cite{lin2024codesign}. These methods are both amenable to using shells, first described in \cite{higgott2021subsystem} as gauge-fixings, and then used in \cite{strikis2023quantum} to prove the existence of thresholds for surface codes under dropouts. In our comparison we do not use shells, but \cite{wei2024low} discusses how they improve performance for the bandage method, and we expect the same would hold for LUCI. Appendix~\ref{app:shells} discusses how one could implement shells with LUCI.

\begin{figure*}[ht!]
    \centering\includegraphics[width=.97\linewidth]{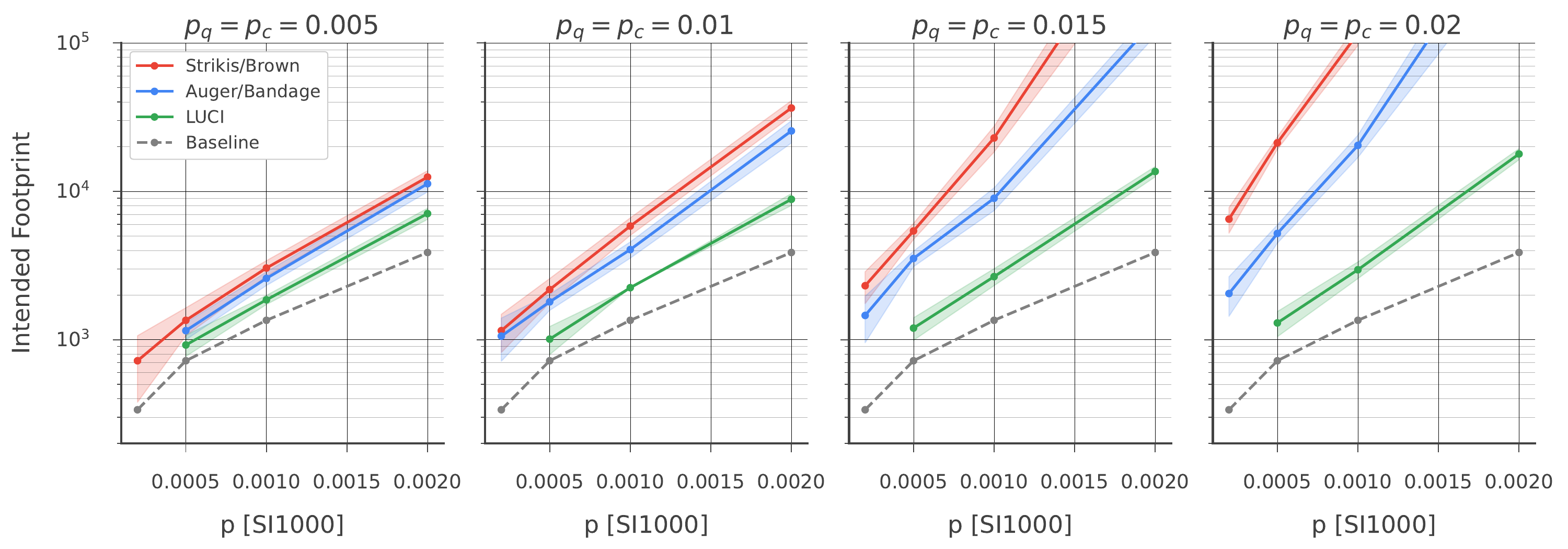}
    \caption{\textbf{Teraquop footprint plot for LUCI. }100 grids were sampled for each error rate and dropout rate at $\ell \in {7, 11, 15}$ and then projected to $10^{-12}$. Error rates are for codeblocks of size $\ell \times \ell \times \ell$ for the dropout-free baseline circuits, $\ell \times \ell \times 2\ell$ for the Strikis/Brown and Auger/Bandage circuits, and $\ell \times \ell \times 4\ell$ for the LUCI circuits. The missing low-error points for LUCI were due to a lack of errors being found for those ensembles, while the missing point on the red curve in the rightmost plot is due to the threshold at that dropout rate being exceeded. Error bars are $5\sigma$ error bars using bootstrapping.}
    \label{fig:teraquop}
\end{figure*}
In Fig.~\ref{fig:example grid simulation}, we compare the three methods on the example grid presented in Fig.~\ref{fig:example grid}. The LUCI circuit used is the one built in Figs.~\ref{fig:LUCI example1} and \ref{fig:LUCI example2}. By looking at the slopes of the three curves, along with the gray reference curve, we can see the impact of distance loss as the error rate improves. The baseline distance-9 circuit has a slope of 5 in the log-log plot, because it takes $\lfloor\frac{d+1}{2}\rfloor$ errors to cause a logical error. When dropouts are added into the device, the distance is reduced to 3 for the Strikis/Brown and Auger/Bandage methods, and 7 for LUCI. The difference between the Strikis/Brown and Auger/Bandage methods is around the broken data qubit and broken coupler labeled as (e) in Fig.~\ref{fig:example grid}. As explained in Ref.~\cite{wei2024low}, this configuration produces a ``bridge'' qubit, which the Auger/Bandage method preserves and the Strikis/Brown method removes. Otherwise the two circuits are very similar, and perform near identically as seen in the figure. LUCI also keeps this qubit, and only removes an additional qubit for the two-broken-coupler and broken-coupler-near-broken-qubit configurations labeled by (d) and (f) in Fig.~\ref{fig:example grid}. At an error rate of 0.001, LUCI shows more than a $30\times$ improvement in logical error rate per round over Auger/Bandage and Strikis/Brown methods for this example grid.

To look at the impacts LUCI has on scalability, in Fig.~\ref{fig:heatmaps} we look at distances and logical error rates as a function of dropout rates for ensembles of grids with a patch diameter of 15. For each pair of qubit and coupler dropout rates (other than the trivial $p_q = p_c = 0$ case) we sampled 250 random grids and built circuits using each method. For distances we took and average of $\text{min}(d_x, d_z)$ across the different grids. All logical error rates were sampled for SI1000(0.001) noise. 

As can be seen in the top set of heatmaps, LUCI far outperforms other methods in terms of distance for all dropout rates considered, improving average distance from 9.1 to 13.1 in the $p_q = p_c = 0.01$ case. This makes sense since LUCI is equal or better than all other method for isolated dropouts in terms of distance preservation, losing no distance for couplers and measurement qubits, and only having distance reduced by 1 in the case of qubit dropout on qubits usually used as data qubits. One key feature to point out is that the LUCI heatmap for distance shows a dramatic asymmetry between coupler and qubit dropouts, while other methods are more symmetric. This is because unlike methods which avoid a broken coupler by removing the attached data qubit, LUCI preserves all qubits and maintains distance. 

In the logical error per round heatmaps on the bottom row, LUCI outperforms the other two methods for every random dropout grid considered. For the same $p_q = p_c = 0.01$ case, the improvement in median logical error rate per round is $25\times$.  The improvements in logical error per round follow from the distance improvements shown in the top set of heatmaps, and become even more significant at lower error rates. Consequently, the asymmetry between couplers and qubits becomes more appreciable as error rates reduce. An intuitive picture for this can be seen by considering the logical fault path, i.e. the logical operator found by adding the maximum likelihood matching from the correct and incorrect logical coset after decoding. This path is rarely exactly length $d$ for error rates near threshold, as the combinatoric coefficient in the logical error polynomial dominates and favors longer strings. However as physical error rates decrease, the penalty for higher-weight strings begins to dominate and logical error rate performance is better described by distance, which is a worst-case metric for a quantum error correcting protocol.

Distributions of distance and logical error rate per round for the 250 random instances with $p_q = p_c = 0.01$ can be found on the right side of the figure. The data confirms that LUCI does appreciably better in terms of both distance and logical error rate, and also has a much tighter distribution. This means that there are far fewer catastrophic dropout arrangements which lead to the heavy tail of negative outliers seen for the Strikis/Brown and Auger/Bandage methods. There is one instance for which the LUCI distance was 4, far outside the distribution. This issue was caused by a corner-case which should have been handled differently in the automated circuit generation, and not by a fundamental issue with the construction.

To further look at how LUCI helps when scaling quantum processors, we provide a set of teraquop footprint plots in Fig.~\ref{fig:teraquop}. A teraquop footprint plot shows curves with logical error rate pinned at $10^{-12}$, hence the name. Along the x-axis we vary the physical error rate in the simulation, while the y-axis shows the physical footprint needed to attain the desired logical error rate. The three dropout methods are shown for $\left(p_q, p_c\right) \in {(0.005, 0.005), (0.01, 0.01), (0.015, 0.015), (0.02, 0.02)}$, along with a baseline dropout-free circuit to act as a reference point. In the plot, we use error rates for a $\ell \times \ell \times \ell$ block for the regular surface code, a $\ell \times \ell \times 2\ell$ block for the Strikis/Brown and Auger/Bandage methods, and a $\ell \times \ell \times 4\ell$ block for the LUCI circuits, to account for the differences in timelike error. Despite this effectively doubling the logical error per round relative to the other methods, at $p_q = p_c = 0.01$ and $p=0.001$ LUCI reduces the needed footprint by more than $25\%$ relative to the Auger/Bandage method. Further optimizations in constructing diagrams are possible to improve the timelike distance, and detailed stability experiments may allow us to shorten the timelike extent of the codeblocks.

\section{Conclusion}\label{sec:conclusion}
In this manuscript we presented LUCI, a framework for building error correction circuits adapted to dropout grids. The main benefits are its flexibility, and the reduction in penalty for broken qubits or couplers. For experimentally reasonable error rates and dropout parameters, we see improvements of $1.4\times$ in distance and $25\times$ in logical error rate per round. This improvement in logical error rate will become more significant as systems improve in both size and performance. The flexibility of the method allows for swapping data and measure qubit roles, as well as dynamically editing QEC circuits to work around TLS's that appear on the device. Our hope is that this reduction in the cost of dropouts enables superconducting qubit hardware research groups to further push the envelope during fabrication.

While we do not discuss logical gates in this work, LUCI circuits should still admit the same operations as the standard surface code. In particular, the boundaries are the same as the usual surface code, so lattice surgery should be a viable option for logical Bell measurements and CNOT operations, but implementation is left to future work. There are also a number of possible performance optimizations, both for generating better LUCI diagrams and integrating known techniques in the field like shells, which are discussed in App.~\ref{app:shells}. We have also seen that ensembled decoders such as Harmony~\cite{shutty2024efficientnearoptimaldecodingsurface} and Libra~\cite{jones2024improvedaccuracydecodingsurface} perform well for LUCI circuits, indicating that there could be value in optimizing decoders for LUCI circuits.

The flexibility provided by LUCI can also be used to build error correction circuits which are anisotropic and aperiodic, or even randomly generated. Along with the fact that LUCI circuits do not measure a consistent end-cycle error correcting code, this indicates that LUCI could be a possible step towards code-free fault-tolerant processes, like \cite{nickerson2018measurementbasedfaulttolerance, Newman2020generatingfault, Hastings2021dynamically, fu2024errorcorrectiondynamicalcodes}. 

\section{Author Contributions}
Dripto and Adam conceived the project based on previous work from Matt and Craig. 
Matt helped with the detecting region and logical operator structure.
Craig worked on a qubit dropout construction which was integrated into the LUCI framework. 
Adam wrote the first iteration of LUCI circuit generation code, Dripto wrote subsequent versions, using tools written by Craig.
Noah helped with decomposing detector error models for LUCI circuits, and generating the Auger/Bandage circuits. 
Dripto wrote the first draft of the manuscript. 
Dripto, Matt, and Noah, revised the manuscript with input from Craig and Adam.

\section{Acknowledgements}
The authors would like to thank Hartmut Neven, the Google Quantum AI team, and Google leadership, for the environment and resources to enable this research. 
The authors would also like to thank Dan Browne, Laleh Aghababaie Beni, Oscar Higgott, Matt Reagor, Dave Bacon, Austin Fowler, Alec Eickbusch, Alexis Morvan, Kevin Satzinger, Hao Tran and Cody Jones, for helpful input on the project and manuscript, and Joe Iverson and Catherine Leroux for pointing out an issue with one of the figures. 


\onecolumngrid
\appendix
\clearpage

\section{SI1000 Noise Model}\label{app:si1000}
The noise model used in this paper is SI1000, a superconducting inspired noise model which assumes a $1000 ns$ cycle time. It is a single parameter noise model with noisy gates defined in Table~\ref{tab:noise_gates} and error rates defined in Table~\ref{tab:noise_model}:

\begin{table}[ht]
    \centering
    \resizebox{\linewidth}{!}{
    \begin{tabular}{|c|l|}
         \hline
         \textbf{Noisy Gate} & \textbf{Definition} \\
         \hline
         $\text{AnyClifford}_2(p)$ & \text{Any two-qubit Clifford gate, followed by a two-qubit depolarizing channel of strength $p$.} \\
         \hline
         $\text{AnyClifford}_1(p)$ & Any one-qubit Clifford gate, followed by a one-qubit depolarizing channel of strength $p$. \\
         \hline
         $\text{R}_{Z}(p)$ & Initialize the qubit as $\ket{0}$, followed by a bitflip channel of strength $p$. \\
         \hline
         $\text{R}_{X}(p)$ & Initialize the qubit as $\ket{+}$, followed by a phaseflip channel of strength $p$. \\
         \hline
         $M_Z(p, q)$ & Measure the qubit in the $Z$-basis, followed by a one-qubit depolarizing channel of strength $p$, \\
         & and flip the value of the classical measurement result with probability $q$.\\
         \hline
         $M_X(p, q)$ & Measure the qubit in the $X$-basis, followed by a one-qubit depolarizing channel of strength $p$, \\
         & and flip the value of the classical measurement result with probability $q$. \\
         \hline
         $M_{PP}(p, q)$ & Measure a Pauli product $PP$ on a pair of qubits, \\
         & followed by a two-qubit depolarizing channel of strength $p$, \\
         & and flip the classically reported measurement value with probability $q$. \\
         \hline
         $\text{Idle}(p)$ & If the qubit is not used in this time step, apply a one-qubit depolarizing channel of strength $p$. \\
         \hline
         $\text{ResonatorIdle}(p)$ & If the qubit is not measured or reset in a time step during which other qubits are \\ &  being measured or reset, apply a one-qubit depolarizing channel of strength $p$. \\
         \hline
    \end{tabular}
    }
    \caption{
        Modified from \protect\cite{gidney2022benchmarking}. Noise channels and the rules used to apply them.
        Noisy rules stack with each other - for example, Idle($p$) and ResonatorIdle($p$) can both apply depolarizing channels in the same time step.
    }
    \label{tab:noise_gates}
\end{table}

\begin{table}[ht]
    \centering
    \begin{tabular}{|c|l|l|}
        \hline
         \textbf{Name}
             & \begin{tabular}{@{}l@{}}Uniform Depolarizing\end{tabular}
             & \begin{tabular}{@{}l@{}}Superconducting Inspired (SI1000)\end{tabular}
        \\\hline
        \textbf{Noisy Gateset}
            &\noindent\begin{tabular}{@{}l@{}}
                $\text{CX}(p)$\\
                $\text{CXSWAP}(p)$\\
                $\text{AnyClifford}_1(p)$\\
                $\text{R}_{Z/X}(p)$\\
                $M_{Z/X}(p, p)$\\
                $M_{PP}(p, p)$\\
                $\text{Idle}(p)$\\
            \end{tabular}
            &\begin{tabular}{@{}l@{}}
                \vspace{-0.25cm}
                {} \\
                $\text{CZ}(p)$\\
                $\text{ISWAP}(p)$\\
                $\text{AnyClifford}_1(p/10)$\\
                $\text{Init}_Z(2p)$\\
                $M_Z(p, 5p)$\\
                $M_{ZZ}(p, 5p)$\\
                $\text{Idle}(p/10)$\\
                $\text{ResonatorIdle}(2p)$\\
            \end{tabular}
        \\\hline
    \end{tabular}
    \caption{
        Modified from \protect\cite{gidney2022benchmarking}. Details of the error models used in this paper.
        See Table~\ref{tab:noise_gates} for definitions of the noisy gates.
    }
    \label{tab:noise_model}
\end{table}

\section{Dropouts near boundaries or clusters}\label{app:boundary dropouts}
\begin{figure}[h]
    \centering  
    \includegraphics[width=0.7\linewidth]{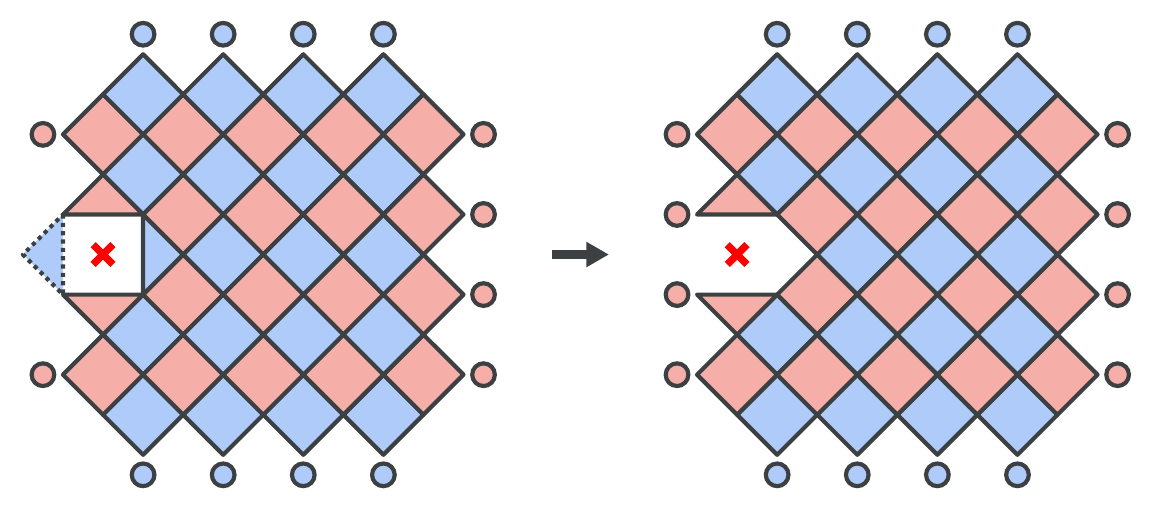}
    \caption{An example situation in which a missing qubit near the boundary can lead to an unpairable gauge operator. The dotted blue triangle indicates where a gauge would have to have existed to pair with the existing Z-type gauge above the missing qubit. Since the dotted gauge is not present, there is no way to pair the Z-type gauge into a super-stabilizer which commutes with its neighbors, so it must be removed.}
    \label{fig:boundary dropout}
\end{figure}
Following the example in Ref~\cite{auger2017fault}, dropouts near the boundary can lead to a gauge which does not have another gauge to join with to form a super-stabilizer, as seen in Fig.~\ref{fig:boundary dropout}. In this case, there ends up being no solution to the pairing problem which includes this gauge operator, and it must be removed from the circuit. Once removed, the adjacent gauge operators can be individually promoted to stabilizers, as there no longer is an anticommuting operator present in the gauge group. This leads to distance only reducing in one basis. Note that in the example the remaining gauge operators are shown as triangles, which would be measured by L-shapes, but a possibly more performant solution would be to use I-shapes instead.

\begin{figure}
    \centering
    \includegraphics[width=0.35\linewidth]{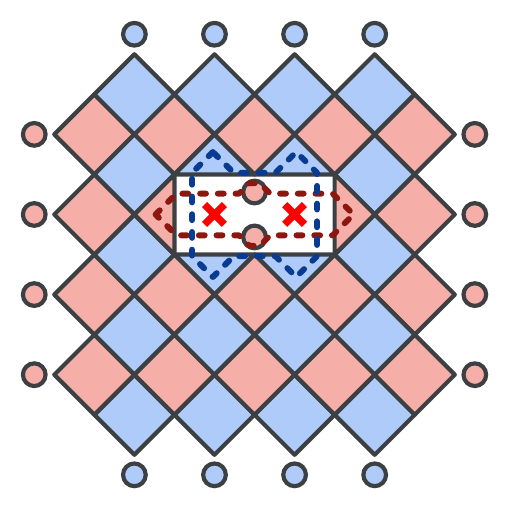}
    \caption{\textbf{Example of a clustered horizontal dropout configuration. }Unlike a case where the middle two weight-1 X-type gauge operators are measured with a shared measure qubit, the four Z-type gauges are all combined into a single mid-cycle superstabilizer.}
    \label{fig:spikespike}
\end{figure}
Another case where gauge operators end up tricky to pair is when there are multiple qubits broken near each other. In \cite{auger2017fault} and \cite{wei2024low}, they handle the case of diagonal broken data qubits by using so-called ``bridge'' qubits, where a measure qubit is used to measure a weight-2 gauge operator on diagonal data qubits. This weight-2 gauge commutes with the pairs of gauges on either side, allowing for two separate super-stabilizers of the opposite basis. In the mid-cycle picture LUCI operates in, measurements are done in place, so this disjoint weight-2 gauge operator would be split into two weight-1 measurements, as showing in Fig.~\ref{fig:spikespike}. To make the commutations work out, the two super-stabilizers of the opposite type which touch the weight-1 gauge operators must be merged, unlike in the end-cycle case.

\section{Detector slices for LUCI example}\label{app:detslices}
\begin{figure}[h]
    \centering
    \includegraphics[width=0.95\linewidth]{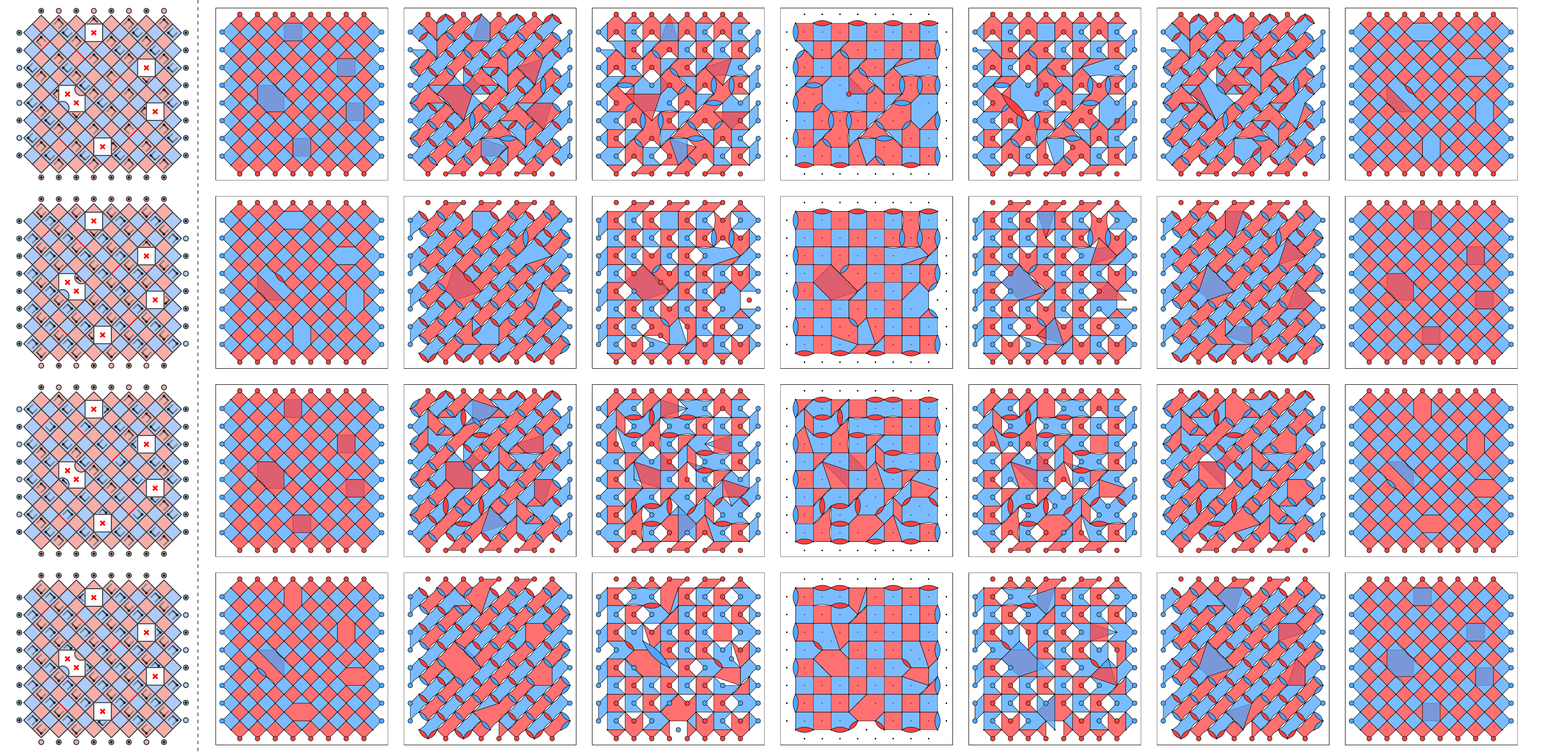}
    \caption{Timelike slices of the detecting regions in a LUCI circuit over layers of the circuit, with the corresponding part of the diagram shown on the left. Each round starts and ends at the mid-cycle state, with some mid-cycle stabilizers and gauge operators being measured as indicated by the shapes. For simplicity we do not include the gate operations in the diagram.}
    \label{fig:detslice}
\end{figure}

In Fig.~\ref{fig:detslice} we show timelike slices of the detectors in four consecutive bulk rounds of a LUCI circuit, along with the diagrams for each round. Each round starts at the midcycle, and is then modified by two layers of CNOT layers. This folds some of the detecting regions into single qubit operators, which are then measured. The qubits which are measured are then reset, causing a new detecting region to open in the second half of the round, until the state is returned to the shared mid-cycle state for the next round to pick up. This detector slice diagram emphasizes the fact that LUCI circuits do not implement a specific end-cycle code, as the cross-sections at the measurement layer (the fourth column of slices) are different in each round. Instead, the LUCI circuits are well behaved at the mid-cycle layer, where the handoff occurs between subsequent rounds.

\section{Using Shells with LUCI}\label{app:shells}
In the appendix of \cite{wei2024low} they discuss how their technique benefits from shells, a technique for dropouts first described in Ref.~\cite{strikis2023quantum} which uses the schedule-induced gauge fixing technique introduced in Ref.~\cite{higgott2021subsystem}. The basic idea is that by measuring gauge operators of the same type some number of times before switching bases, one can treat the gauge operators like stabilizers in the successive repetitions, with the super-stabilizer only being used in the first round of a given basis, as the individual gauge operators would be scrambled by the measurements of the opposite basis. In Ref.~\cite{strikis2023quantum} they repeat the gauge operators around the dropout a number of times proportional to the patch diameter of the dropout region, while in Ref.~\cite{wei2024low} they use global X and Z layers are study different ways to weight the patch diameters in the circuit. 

These ideas are all applicable to LUCI circuits as well, where the global strategy involves simply repeating the first two rounds of the diagram a number of times, then the second pair of rounds, using whatever weighting desired. You could also make mixed-basis rounds to repeat larger dropout regions more than smaller ones. We believe that the impact of shell methods would be smaller in the case of LUCI than in other methods, as the technique shines most when there are large super-stabilizers containing many gauge operators, and LUCI tends to produce smaller super-stabilizers than other techniques in most cases, however testing this hypothesis is left to future work.
\end{document}